\documentclass[prd, twocolumn, nofootinbib, floatfix]{revtex4}

\usepackage{epsfig} 
\usepackage{amsmath}
\usepackage{amsbsy,color}

\newcommand{\beq}{\begin{equation}}
\newcommand{\eeq}{\end{equation}}
\newcommand{\beqa}{\begin{eqnarray}}
\newcommand{\eeqa}{\end{eqnarray}}
\newcommand{\om}{\Omega_m}
\newcommand{\omw}{\Omega_w}
\newcommand{\ow}{\Omega_w}

\newcommand{\wtot}{w_{\rm tot}}
\newcommand{\ac}{a_{\rm acc}}
\newcommand{\zac}{z_{\rm acc}}
\newcommand{\nacc}{N_{\rm acc}}

\begin{document} 

\title{Dark Before Light: Testing the Cosmic Expansion History through 
the Cosmic Microwave Background} 
\author{Eric V.\ Linder$^{1,2}$ and Tristan L.\ Smith$^1$} 
\affiliation{$^1$Berkeley Center for Cosmological Physics \& Berkeley Lab, 
University of California, Berkeley, CA 94720, USA \\ 
$^2$Institute for the Early Universe, Ewha Womans University, Seoul, Korea} 
\date{\today}

\begin{abstract} 
The cosmic expansion history proceeds in broad terms from a radiation 
dominated epoch to matter domination to an accelerated, dark 
energy dominated epoch.  We investigate whether intermittent periods 
of acceleration are possible in the early universe -- between Big Bang 
nucleosynthesis (BBN) and recombination and beyond. We establish that 
the standard picture is remarkably robust: 
observations of anisotropies in the cosmic microwave background exclude 
any extra period of accelerated expansion 
between $1 \leq z \lesssim 10^5$ (corresponding to 
$5\times10^{-4}\ {\rm eV} \leq T \lesssim 25\ {\rm eV}$).  
\end{abstract} 

\maketitle

\section{Introduction \label{sec:intro}}

Remarkably little is known about the detailed expansion history of 
the universe.  While it proceeds from an epoch of radiation domination, 
to matter domination, both decelerating, to recent acceleration, even 
in the best constrained periods -- Big Bang nucleosynthesis and 
recombination -- our specific knowledge of the scale factor as a function 
of time, $a(t)$, is modest.  Limits exist on 
the magnitude of the expansion rate, or Hubble parameter $H=d\ln a/dt$, 
to $\sim5\%$ in the best cases \cite{kaplinghat,zahn}, but not on its 
behavior, for example its 
slope $d\ln H/d\ln a$.  The latter quantity is directly related to the 
total equation of state $\wtot$ of the energy contents and governs whether 
the deceleration parameter $q=-1-d\ln H/d\ln a=(1+3\wtot)/2$ is positive 
(decelerating expansion) 
or negative (accelerating expansion).  The uncertainties on $q$ can 
easily be of order unity -- hence allowing either sign -- for any short 
period (such as Big Bang nucleosynthesis or recombination) in the early 
universe. 

This article addresses this lacuna in our understanding through 
investigation of the effects of a period of early acceleration on 
the cosmic microwave background (CMB) anisotropies.  This is complementary 
to the recent analysis of Ref.~\cite{uniq} that put severe limits on 
acceleration at redshifts $z\approx2$-1000 from effects on matter 
perturbation 
growth. 

While knowledge of the detailed expansion history is useful in itself, 
constraints on an early period of acceleration also impact any attempt to 
ameliorate the coincidence problem of why acceleration is occurring now 
by adopting models in which acceleration is a persistent or occasional 
phenomenon (e.g.\ \cite{dodelson00,griest,barenboim}).  Sufficiently tight 
limits could force us to focus on 
dynamical mechanisms for causing recent acceleration, or to accept the 
coincidence. 

Other rationales for exploring the expansion history for an additional 
period of acceleration, generally 
at much earlier epochs than we consider, include: thermal inflation 
models, which can solve the moduli problem and impact baryogenesis 
\cite{lyth}, secondary inflation, which can affect the scalar 
perturbation tilt and matter power spectrum \cite{ackerman,liddle}, 
and multiple inflation, which can put features in the curvature perturbation 
power spectrum \cite{sarkar}, and scalar torsion cosmology \cite{shie}. 

In Sec.~\ref{sec:method} we present the basic characteristics of early 
acceleration and outline the modifications to the background and 
perturbation equations necessary for determining the CMB anisotropy 
spectrum.  The physical impact of early acceleration on the CMB is 
analyzed in Sec.~\ref{sec:cmb} and the numerical results computed. 
We conclude with a discussion of the constraints on the duration of 
such an accelerating expansion in Sec.~\ref{sec:sumy}.

%%%%%%%%%%%%%%%%%%%%%%%%%%%%%%%%%%%%%%%%%%%%%%%%%%%%%%%%%%%%%%%
\section{Adding Early Acceleration \label{sec:method}}

As discussed in Ref.~\cite{uniq}, there exist two ways of obtaining 
an intermittent period of acceleration: driving the dark energy 
of state highly negative, $w\ll-1$, called superacceleration, 
or having a relatively high dark energy density with a fairly 
negative equation of state $w<-1/3$, but then $w$ must go positive 
in order to reduce the dark energy density and permit matter or 
radiation to dominate again, called superdeceleration.  The first 
approach was shown not to be viable, especially at high redshifts, 
due to the intrinsic dynamics of the dark energy density evolution 
for supernegative equations of state, so we only consider superdeceleration 
models. 

%%%%%%%%%%%%%%%%%%%%%%%%%%%%%%%%%%%%%%%%%% 
\subsection{Acceleration Model} 

To model the dark energy, we take a $\Lambda$CDM universe that 
undergoes a finite period of deviation from $w=-1$.  The e-fold model 
of Ref.~\cite{linhut05} provides a useful parametrization of such a 
transition, 
with one step up to positive $w$ and then a later step back down to 
$-1$.  This e-fold Ansatz has the physical advantage of defining a 
characteristic scale for the rapidity of the transition, in terms of 
number of e-folds of expansion, and the mathematical advantage of giving an 
analytic expression for the Hubble parameter.  

The dark energy equation of state as a function of scale factor $a$ is 
\beq 
w(a)=-1+\frac{w_j+1}{1+(a/a_{d})^{1/\tau}}-
\frac{w_j+1}{1+(a/a_{t})^{1/\tau}} \,, 
\eeq 
where $w_j$ is dark energy equation of state during superdeceleration, 
$a_{t}$ is the scale factor when $w$ is in the middle of the jump up, 
$a_{d}=a_{t}\,e^N$ is the scale factor when $w$ is in the middle of 
the jump back down, $1/\tau$ is the rapidity of the jumps, i.e.\ 
$dw/d\ln a(a_t)=(w_j+1)/(4\tau)$, and $N$ measures the number of e-folds 
between the transitions.  For small $\tau$, the transition approaches a 
step function. 

\begin{figure}
  \begin{center}{
  \includegraphics[width=\columnwidth]{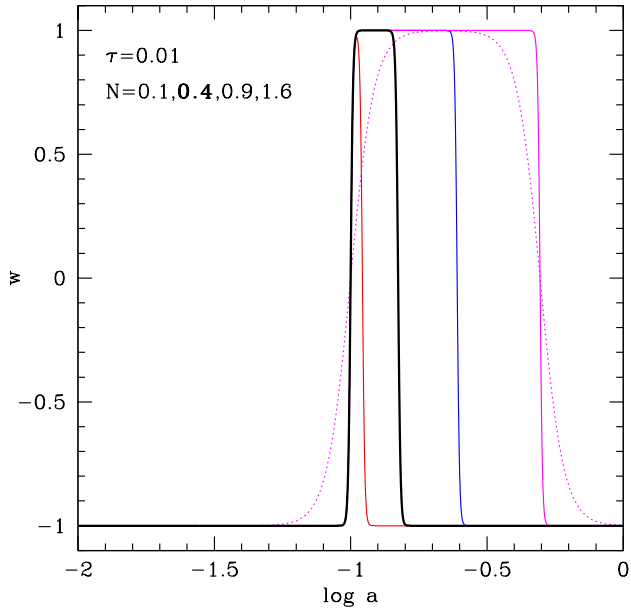}
  }
  \end{center}
  \caption{The double e-fold form for $w(a)$ has useful physical 
and mathematical properties to describe a step transition.  For 
inverse rapidity $\tau\ll1$, the transition approximates a step, and for 
width $N\gtrsim 10\tau$ the value at the top of the step stays 
constant for $N$ e-folds.  Solid curves all take $\tau=0.01$ and 
show the behaviors for different values of $N$; the dotted, magenta 
curve takes $N=1.6$ but for $\tau=0.1$ to show the impact of the 
rapidity.  All curves take the transition up at $a_{t}=0.1$. 
}
  \label{fig:efoldw}
\end{figure}

At times much earlier than the transition up, $w(a\ll a_{t})=-1$, 
and at times much later than the transition back down, $w(a\gg a_{d})=-1$, 
while if the transitions are well separated then 
$w(a_{t}\ll a\ll a_{d})=w_j$.  A true leveling off at $w_j$ is achieved 
for $N\gtrsim 10\tau$.  These behaviors are illustrated in 
Fig.~\ref{fig:efoldw}.

The dark energy density evolution follows 
\beq 
\rho_w(a)=\rho_{w,0}\,\left(\frac{1+a_{t}^{1/\tau}}{1+a_{d}^{1/\tau}}\right)^{3\tau\Delta w}\, 
\left[\frac{1+(a/a_{d})^{-1/\tau}}{1+(a/a_{t})^{-1/\tau}}\right]^{3\tau\Delta w}\,, 
\eeq 
where $\Delta w=w_j+1$.  The dark energy acts like a cosmological 
constant from the present back to $a\approx a_{d}$, with energy density 
$\rho_{w,0}$, then climbs during the step and levels off to a constant 
$\rho_{w,0}\,e^{3N\Delta w}$ for $a\lesssim a_{t}$.  Viewed the other 
way around, the early dark energy density acts like a larger cosmological 
constant than usual, allowing early acceleration, then the period of 
superdeceleration dilutes the density down to the level matching the 
value today. 

It is useful to transform into variables more closely related to the 
expansion physics such as the scale factor $a_{\rm acc}$ when the 
accelerated expansion starts (determined by the condition 
$\wtot=\sum w_i(a_{\rm acc})\,\Omega_i(a_{\rm acc})=-1/3$ where the 
sum runs over all components) and the 
number of e-folds of accelerated expansion, $N_{\rm acc}$.  Using these 
as the fundamental parameters, one has 
\begin{eqnarray}
a_t &=& a_{\rm acc} e^{N_{\rm acc}} \\
a_d &=& a_{\rm acc} e^{N_{\rm acc}} 
\left[\frac{\Omega_B(1+3w_B)}{2\ow a_{\rm acc}^{3(1+w_B)}}\right]^{1/[3(1+w_j)]} \\ 
N&=&\frac{-(1+w_B)}{1+w_j}\ln\ac-\frac{1}{3(1+w_j)} 
\ln\left[\frac{2\ow}{\Omega_B(1+3w_B)}\right] \,, 
\end{eqnarray}
where $N$ is the number of e-folds of the superdeceleration period, 
$\Omega_i$ is the fractional energy density in component $i$, and a 
subscript $B$ denotes the dominant component excepting dark 
energy (e.g.\ $w_B=1/3$ when radiation dominates over matter, $w_B=0$ 
when matter dominates over radiation). 

The necessary condition to have any acceleration is 
\beq 
N>\frac{1}{3(1+w_j)}\,\ln\left[\frac{\Omega_B(1+3w_B)}{2\omw a_t^{3(1+w_B)}}\right]\,. 
\eeq 
For prerecombination acceleration, $a_t$ will be very small and one 
would need more than 3 e-folds of superdeceleration to allow any 
acceleration at all.  Generally the number of e-folds of superdeceleration 
is much larger than the number of e-folds of acceleration; it is this that 
imposes strong constraints on early acceleration from effects on the CMB. 

Of course we need not restrict ourselves purely to testing for early 
acceleration.  A modification of the expansion history that is 
insufficient to give acceleration, i.e.\ where the early dark energy density 
is not quite high enough and $\nacc<0$, can still affect CMB observations. 
We therefore will also be interested in the deviation from the standard 
expansion history 
\beq 
R=\frac{\Delta H^2}{H^2_{\rm std}}\approx\frac{\rho_w(a)}{\rho_B(a)}= 
\frac{1+3w_B}{2}\left(\frac{a}{\ac}\right)^{3(1+w_B)}r(a)\,,
\eeq 
where again $B$ denotes the main background component (radiation or 
matter), the second equality holds when one component dominates 
in the standard scenario, and $r(a<a_t)=1$, 
$r(a_t<a<a_d)=(a/a_t)^{-3\Delta w}$.  We can ask when $R$ exceeds some value, 
$R_*=0.1$ or 1 say.  The number of e-folds $N_*$ when $R>R_*$, and the 
maximum deviation $R_{\rm max}$, are given by 
\beqa 
N_*&=&\frac{1+w_j}{3(1+w_B)(w_B-w_j)}\,\ln\frac{R_*}{R_{\rm max}} \label{eq:nstar}\\ 
R_{\rm max}&=& 
\frac{1+3w_B}{2}\,e^{3\nacc(1+w_B)} \,. \label{eq:rmax} 
\eeqa 

In the analysis we take $w_j=+1$, the maximum value consistent with a 
canonical 
scalar field, which gives the most conservative case in that this 
permits the shortest period of superdeceleration for a given acceleration. 
For the same reason we adopt $\tau\ll1$.  Therefore 
there are two free parameters: one describing the time when the transition 
occurs, and one for its duration.  
These determine the physical effects on the CMB anisotropy power spectrum.

%%%%%%%%%%%%%%%%%%%%%%%%%%%%%%%%%%%%%%%
\subsection{Perturbation Equations} 

When modifying the expansion history through an additional, dark 
component one must also account for modifications to the perturbed 
quantities in order to ensure energy momentum conservation.  We 
include the generalized dark fluid following Ref.~\cite{hu}, taking 
it to be a minimally coupled, canonical scalar field. 
Then the density and velocity perturbations to 
this fluid in synchronous gauge, for each Fourier mode $k$, take the form 
\begin{eqnarray}
\frac{\delta_g'}{1+w} &=& -\frac{\mathcal{H} (1 - w)}{k(1+w)}\delta_g \\
&-& \left[1+9 \frac{\mathcal{H}^2}{k^2}(1 - w) + 3 \frac{\mathcal{H} w'}{k(1+w)} \right] v_g - \frac{1}{2} h',\nonumber\\
v_g' &=& -\frac{\mathcal{H}}{k} v_g + \frac{1}{1+w} 
\left[\delta_g + 3 \frac{\mathcal{H}}{k} (1+w) v_g\right],
\end{eqnarray}
where $h$ is one of the metric perturbations in synchronous gauge 
(see, e.g., \cite{ma_and_bert}), 
$\mathcal{H} = a^{-1} d a/d\eta$, $\delta_g$ is the perturbed energy 
density of the dark fluid, $v_g$ is the velocity perturbation, and the 
prime indicates a derivative 
with respect to $k\eta$ where $\eta$ is the conformal time. 

The metric perturbation $h$ couples these equations to the matter and 
radiation perturbations.  We take adiabatic initial conditions for the 
dark fluid.  Note the presence of the $w'$ term; we have checked 
explicitly that the sharp, but brief, transitions we use do not affect 
the results.  As we vary the inverse rapidity from $\tau=0.05$ 
to 0.02 or 0.1, say, the resulting power spectra remained robust. 
We coded CAMB \cite{camb} with the modified background expansion and 
modified perturbation equations and investigated the effects on the CMB 
power spectrum.

%%%%%%%%%%%%%%%%%%%%%%%%%%%%%%%%%%%%%%%%%%%%%%%%%%%%%%%%%%%%%%%
\section{CMB Power Spectrum and Acceleration \label{sec:cmb}}

There are three main cases for the impact of early acceleration. 
First we will consider the effects of a modified expansion history when 
it occurs after recombination (similar to Ref.~\cite{uniq} but here 
with respect to the CMB rather than matter growth).  In this case 
the dominant effect on the CMB power spectrum will be a shift in the 
angular diameter distance to the surface of last scattering (leading to a 
shift in the location of the acoustic peaks in the power spectrum) as well 
as an extra bump at lower multipoles due to a modified integrated 
Sachs-Wolfe effect.  Second we will consider the effect of a 
modification occurring before recombination.  In this case the evolution 
of the photon perturbations is altered, leading to an enhancement of power 
on small scales.  The third case is when the altered expansion history 
significantly overlaps the recombination period, and is basically a 
superposition of the first two cases.

%%%%%%%%%%%%%%%%%%%%%%%%%%%%%%%%%%%%%%%%%%%%
\subsection{General Properties \label{sec:genprop}} 

Before exploring the individual cases, let us discuss some general 
influences of modified expansion on the CMB.  First, consider the range 
of redshift that 
is accessible through observations of the CMB.  A measurement of  the CMB 
anisotropy for a given multipole $\ell$ corresponds to a physical wave-number 
$k$ through the scaling $k\eta_0\sim\ell$, where $\eta_0\sim10^4\ {\rm Mpc}$ 
is the conformal time today.  A modified expansion history will affect most 
significantly the evolution of the perturbations for modes that are within 
the horizon during the modified expansion.  Since modes are within the 
horizon when $k \eta(z) \gtrsim 1$, we expect that only those multipoles 
with $\ell \gtrsim \eta_0/\eta(z_{\rm mod})$ will be significantly modified.  
Given that observations of the primordial CMB can only be made up to 
$\ell_{\rm max} \sim 3000$ before secondary anisotropies become 
significant, by using the CMB we can expect to be sensitive to modifications 
in the expansion history up to $z_{\rm max} \sim 10^5$.  As we will see, 
modifications to the expansion history at larger redshifts will have a 
diminishing effect on the CMB anisotropies. 

\begin{figure}
  \begin{center}{
  \includegraphics[width=\columnwidth]{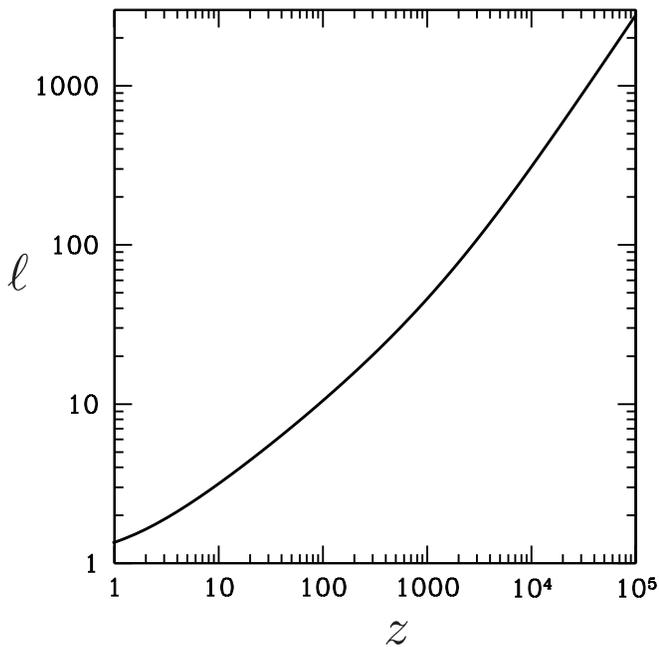}
  }
  \end{center}
  \caption{The minimum multipole affected by modified expansion is plotted 
as a function of redshift in the standard $\Lambda$CDM cosmology.  Only 
multipoles above this had associated wavenumbers $k$ within the horizon 
at redshift $z$.  For measurements of the primary CMB anisotropy up to 
$\ell_{\rm max} \sim 3000$, say, one can therefore probe the expansion 
history out to $z_{\rm max} \sim 10^5$.  
}  
  \label{fig:ellz}
\end{figure}

Figure~\ref{fig:ellz} shows the rough range of multipoles over which 
modifications will show up in the CMB power spectrum.  In fact, effects 
can appear at somewhat smaller multipoles as well since the enhanced energy 
density and hence Hubble parameter persists to later times, i.e.\ 
smaller redshifts, extending the multipole range for fixed $z$ or 
extending the redshift sensitivity for fixed $\ell$.  From 
Eqs.~\ref{eq:nstar} and \ref{eq:rmax} one finds that for $\nacc=0.1$ 
and $R_*=0.5$, say, the effective redshift is lowered by a factor 2, 
hence a modification starting at $z=10^5$ actually affects down to 
$\ell\approx1000$, or conversely sensitivity to $\ell=3000$ actually extends 
to redshifts $z\approx 2\times10^5$.  For $\nacc=1$, the maximum 
redshift probed becomes $z\approx 3\times10^6$.

%%%%%%%%%%%%%%%%

Another generic effect is on the geometric information in the CMB 
from the distance scales involved.  
The location of the acoustic peaks of the CMB anisotropies is determined 
by the simple relation $k_A = \pi/r_s(\eta_{\rm rec})$ 
\cite{husugi} where $r_s$ is the sound horizon.  (We discuss phase 
factors from potential driving in Sec.~\ref{sec:prerec}; they are 
unimportant for the following scaling argument.)  
We can approximately write $r_s \sim \eta_{\rm rec}$, where $\eta_{\rm rec}$ 
is the conformal time at recombination, so that the location of the peaks 
in harmonic space scales as $\ell_p \propto \eta_0/\eta_{\rm rec}$.  Since 
$\eta$ is inversely proportional to $H$ then an increase in $H$ (e.g.\ due to 
the extra dark energy density needed for early acceleration) leads to a 
decrease in $\eta$.  If $H$ is increased after recombination then $\eta_0$ 
decreases whereas $\eta_{\rm rec}$ remains unchanged and the peaks move to 
smaller $\ell$; if $H$ is increased before recombination then 
$\eta_{\rm rec}$ is decreased whereas $\eta_0$ remains nearly unchanged 
(since it receives most of its contribution from late times), leading to 
a shift in the peaks towards larger values of $\ell$. 

Finally, for the models that we consider the redshift of recombination 
$z_{\rm rec}\approx1090$ 
is left nearly unchanged.  We can understand this by noting the redshift 
of recombination is approximately determined when the expansion rate is 
equal to the interaction rate between photons and free electrons, 
\begin{equation}
X_e n_b \sigma_T = H \,,
\end{equation} 
where $n_b$ is the baryon number density, $X_e$ the ionization fraction, 
and $\sigma_T$ the Thomson scattering cross section.  
If $H$ is larger at a given redshift $z$ during recombination then due 
to the detailed balance equations the free electron fraction will similarly 
be larger (the faster the universe is expanding at a given redshift the 
harder it will be for free electrons to combine with protons).  This 
balance (described in more detail in \cite{joneswyse} for the standard 
model and in \cite{lin97} for general $H$) leads to 
$z_{\rm rec}$ remaining nearly unchanged.

%%%%%%%%%%%%%%%%%%%%%%%%%%%%%%%%%%%%%%%%%%%%%%%%%%%%%%%%%%%%%%%
\subsection{Post-Recombination Effects \label{sec:postrec}}

Modifying the expansion history in the post-recombination phase produces 
two distinct physical effects on the CMB temperature power spectrum.  
Both arise from the change in expansion history rather than the photon 
perturbation per se.  
One physical effect is that modifying the expansion history changes the 
angular diameter distance to the surface of last scattering, leading to 
a shift in the location of the acoustic peaks in the power spectrum, 
as already discussed. 
Note that because the modification is strictly post-recombination, no 
change occurs in the sound horizon length, so the shift is given purely 
by the angular distance change.  

For a sharp transition lasting a small 
fraction of an e-fold, one can calculate the effect analytically: 
\beq 
\frac{\delta d_{\rm lss}}{d_{\rm lss}}\approx -\frac{3}{7}\frac{\om}{\omw} 
\frac{1}{\sqrt{\om}d_\Lambda}\,a_d^{7/2} N(1+w_j)\,, \label{eq:ddlss}
\eeq 
where $d_\Lambda$ is the distance to CMB last scattering in the standard, 
no transition case.  The main effect is from the high dark energy density 
before the transition, rather than the transition (period with $w_j=1$) 
itself.  Note that for small $N$ the key parameter, other than the timing 
of the transition $a_d$, is an ``equivalent width'' $N(1+w_j)$ or area of 
the deviation from $w=-1$ in $\ln a$.  For large $N$, the shift increases 
more rapidly than in Eq.~(\ref{eq:ddlss}).  

\begin{figure}
  \begin{center}{
  \includegraphics[width=\columnwidth]{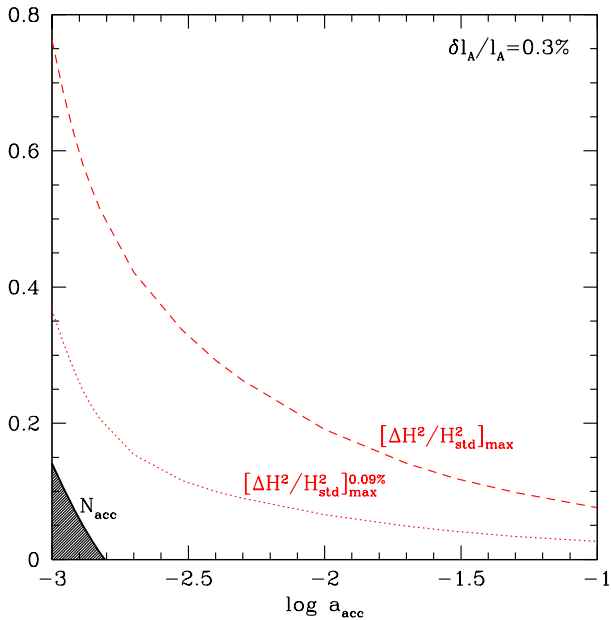}
  }
  \end{center}
  \caption{An upper limit on the number of e-folds of early acceleration 
in the post-recombination case can be derived by requiring that the 
acoustic peak location $l_A$ be shifted by less than 0.3\% (current WMAP 
accuracy) by effects on the distance to last scattering $d_{\rm lss}$.  
This leaves only the tiny shaded corner of parameter space, with 
$\nacc<0.14$ at $z\approx1000$.  Alternately, one could ask how large 
a deviation is allowed from the standard Hubble parameter (squared).  
This is restricted at the tens of percent level (red dashed curve), or 
a factor of 2-3 less for 0.09\% or Planck level constraints on $l_A$ 
(red dotted curve).  In fact, constraints are even more severe if one 
takes into account ISW and other CMB power spectrum information, 
although they loosen if one marginalizes over recent universe 
parameters such as $\om$ (see Fig.~\ref{fig:fisher} for limits taking 
both these effects into account). 
}
  \label{fig:dLSS}
\end{figure}

Observations made by WMAP \cite{wmap} have constrained the location of the 
first acoustic peak to within $0.3\%$.  As a first look, nearly analytically, 
we can use this measurement to exclude regions of the $\nacc$-$\ac$ parameter 
space, fixing all recent 
universe parameters and testing for post-recombination expansion 
modifications.  As seen in Fig.~\ref{fig:dLSS}, almost no early acceleration 
is permitted, except for a tiny region near $z\approx10^3$ amounting to less 
than 0.14 e-folds of acceleration.  The Planck satellite determination of 
the acoustic peak location to $0.09\%$ would completely rule out early 
acceleration (recall that there is a modification of the Hubble parameter 
due to excess energy density even if it does not rise to the level of 
causing acceleration).  Apart from $\nacc$, one can also directly limit 
the modification to the expansion as given by the dashed (WMAP level) 
and dotted (Planck level) curves in Fig.~\ref{fig:dLSS}. 
We revisit the constraints rigorously in Sec.~\ref{sec:fisher}.

%%%%%%%%%%%%%%%%%%%%%%%%

The second effect is that the change in the expansion rate leads to the 
decay of the gravitational potentials, giving rise to a modified 
integrated Sachs-Wolfe (ISW) effect.  The location of the extra ISW power is 
related to the time of the expansion modification, with earlier 
transitions leading to an effect at higher multipoles (where cosmic 
variance is not as severe).  

\begin{figure*}
  \begin{center}{
  \includegraphics[width=\textwidth]{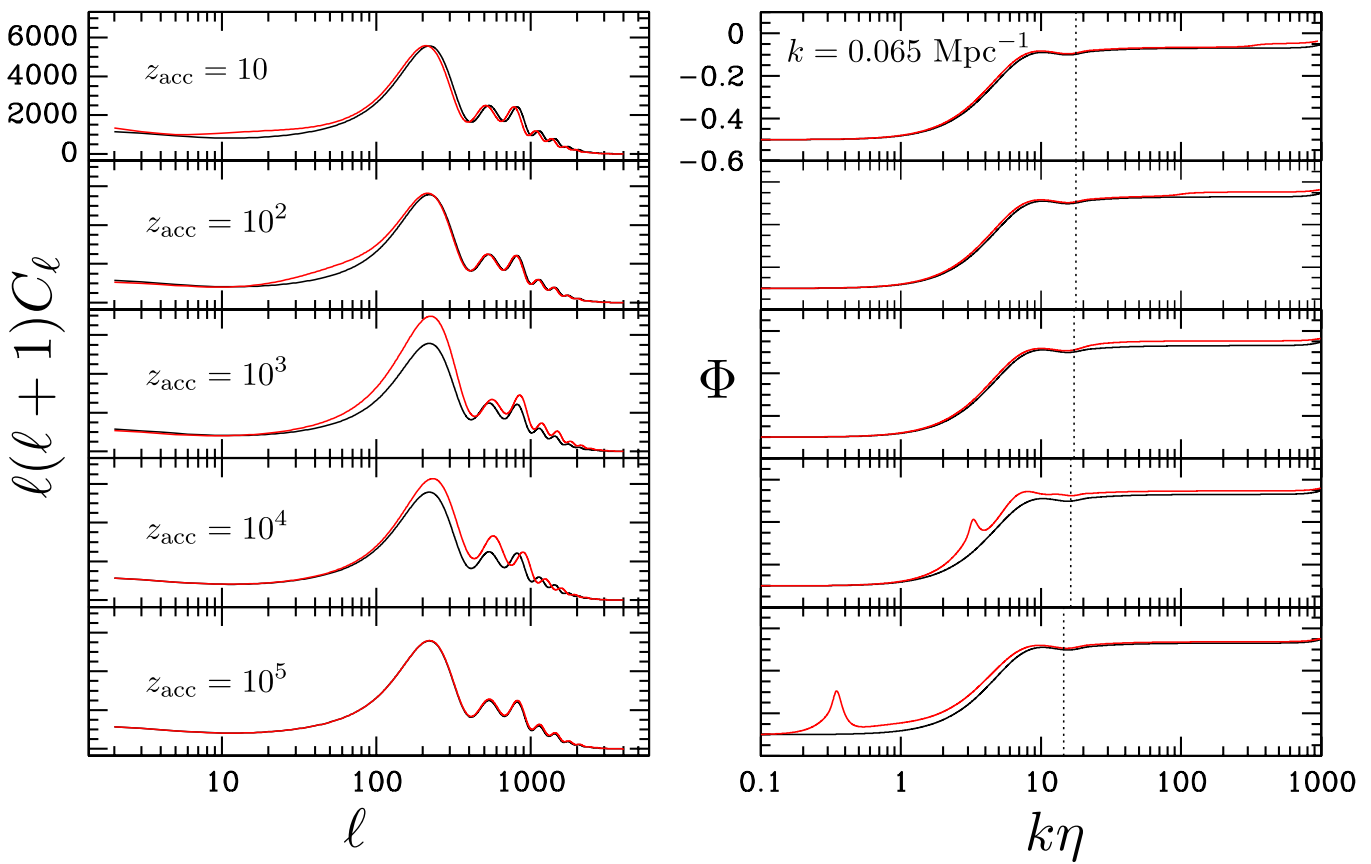}
  }
  \end{center}
  \caption{\emph{Left}: The CMB temperature power spectrum is plotted 
for the standard expansion history (black curves) and a modified expansion 
history (light, red curves).  In all panels a period of accelerated 
expansion starts at redshift $z_{\rm acc}$ and lasts for 0.1 e-folds.   
\emph{Right}: The evolution of the curvature potential corresponding to 
each case in the left panels is shown for the mode with 
$k=0.065\ {\rm Mpc}^{-1}$.  The vertical dotted lines show the value of 
$k\eta_{\rm rec}$, and one sees how the potential decay occurs 
post-recombination for the top panels and pre-recombination for the 
bottom panels.  For $z_{\rm acc} < z_{\rm rec}$ the power spectrum 
peaks are shifted slightly towards lower $\ell$ since the angular diameter 
distance to the surface of last scattering decreases.  In addition to this 
the modified expansion history causes an evolution in the potentials, as seen 
on the right, leading to an additional integrated Sachs-Wolfe 
bump in the power spectrum (most visible in the $\zac=10$ case).  As 
discussed further in the text, when $z_{\rm acc}\gtrsim z_{\rm dec}$ two 
different effects are important.  The sound horizon at decoupling is 
decreased leading to a shift of the peaks to higher $\ell$ (most visible 
in the $z_{\rm acc} = 10^4$ case).  In addition to this, the right panels 
show that the amplitude of the potential is decreased, which leads to a 
corresponding increase in the amplitude of the photon perturbations seen 
in the left panels. 
} 
  \label{fig:prerec_ps}
\end{figure*}

Figure~\ref{fig:prerec_ps} shows the changes induced in the CMB power 
spectrum from a period of early acceleration lasting for $\nacc=0.1$ e-folds, 
with the top panels corresponding to the post-recombination transition 
discussed in this section.  One can clearly see the rise in ISW power at 
low multipoles (with $\ell$ increasing as $\zac$ does).  The geometric 
shift in the acoustic peak locations to smaller multipoles is also 
visible, and this shift gets reversed to higher multipoles for 
transition pre-recombination.  We discuss the other effects from 
pre-recombination acceleration, and the right panels displaying the 
potential perturbations, in the next subsection.

%%%%%%%%%%%%%%%%%%%%%%%%%%%%% 

%%%%%%%%%%%%%%%%%%%%%%%%%%%%%%%%%%%%%%%%%%%%%%%%%%%%%%%%%%%%%%
\subsection{Pre-Recombination effects \label{sec:prerec}}

Modifying the expansion history pre-recombination leads to a modification 
of the observed power spectrum on small scales, due mainly to the influence 
on the photon perturbations between when a given wavemode 
enters the horizon and recombination.  There is also a geometric shift 
but since the sound horizon length gets its main contribution from near 
recombination, a sufficiently early transition has little 
effect on the acoustic scale (the shift goes roughly as $a_t N(1+w_j)$ for 
$N\ll1$).  

Heuristically, one can view the enhancement of power on small scales due 
to the decay of the potentials in terms of a compressed photon fluid 
rebounding while feeling a reduced gravitational restoring force 
\cite{husugi}.  We 
briefly give a more detailed explanation now. 

While the photons are tightly coupled to the baryons they follow a driven 
harmonic oscillator equation \cite{husugi}
\begin{equation}
\left\{\frac{d^2}{dx^2} + 1\right\} [\delta_{\gamma} + \Phi] = 2 \Phi, 
\end{equation}
where $x \equiv k\eta/\sqrt{3}$, $\Phi$ is the potential, 
and we have neglected the time dependence 
of the sound speed of the photon-baryon fluid and neglected anisotropic 
stress and dropped the damping terms since they are negligible to a 
first approximation.  
The solution which corresponds to adiabatic initial conditions gives 
\begin{eqnarray}
\delta_\gamma(x) + \Phi(x) &=& [\delta_{\gamma}(0)+\Phi(0)] \cos(x) \\
&\quad& +2 \int_0^{x} d x' \Phi(x') \sin[x- x'].\nonumber
\end{eqnarray} 
The modified expansion history, as well as perturbations in the dark fluid 
that driving the modification, leads to a change in the potential 
$\Phi$ (as is well known at late times in $\Lambda$CDM cosmologies).  
This in turn drives the photon perturbations to a larger amplitude.  

The CMB power spectrum for the pre-recombination modification case is 
shown in the bottom panels of Fig.~\ref{fig:prerec_ps}.  For 
$z_{\rm acc} = 10^3$ the modified expansion history leads to an increased 
early ISW which in turn increases the height of the first peak.  The  
heights of the following peaks have increased due to the decrease in the 
amplitude of the potential, shown in the right panel.  The 
$z_{\rm acc} = 10^4$ case clearly shows that with the modified expansion 
history the location of the peaks shift toward higher $\ell$.  As described 
before this is due to the decrease in the sound horizon at decoupling 
because of the increased Hubble parameter.  For modifications at such high 
$z$ there is little effect on the ISW contribution.  

Moving to even higher redshift weakens all the effects since the sound 
horizon scale is mostly governed by conditions near recombination and 
early modifications affect the potential driving of perturbation modes 
at high $\ell$ where they are more strongly damped.   
Observations of the small scale CMB with instruments such as the Atacama 
Cosmology Telescope (ACT) \cite{act} or the South Pole Telescope (SPT) 
\cite{spt} allow us to constrain the expansion history up to $z \sim 10^5$ 
since they measure the primordial anisotropies to $\ell \sim 3000$ 
(see Fig.~\ref{fig:ellz}).  Current constraints allow less than 
0.1 e-folds of accelerated expansion at 
$z_{\rm acc} = 10^5$.  Planck \cite{planck} will be cosmic variance 
limited out to $\ell\approx2500$ and delivers tighter constraints 
as we discuss in Sec.~\ref{sec:fisher}.

%%%%%%%%%%%%%%%%%%%%%%%%%%%%%%%%%%%%%%%%%%%%%%
\subsection{Near-recombination effects \label{sec:nearrec}} 

For a transition near recombination, the effects are basically a 
superposition of the pre-recombination and post-recombination ones, 
since the influence of the extra dark energy density persists from 
before to after recombination. 
One other, more minor effect enters on very small scales.  
The diffusion of photons leads to a damping of perturbations 
$\sim e^{-\ell^2/\ell_D^2}$ below a scale $\ell_D$.  A simple argument 
(see, e.g., \cite{dodelson}) shows that this scale goes as 
\begin{equation}
\ell_D \sim \eta_0 \sqrt{\frac{X_e n_b \sigma_T a}{\eta_{\rm rec}}} \,. 
\end{equation} 
An increased Hubble parameter before recombination will lead to a decrease 
in $\eta_{\rm rec}$, an increase in $\ell_D$, and hence 
an increase in the amplitude of perturbations on small scales.  
This adds somewhat to the enhancement due to the potential decay. 

%%%%%%%%%%%%%%%%%%%%%%%%%%%%%%%%%%%%%%%%%%%%%%
\subsection{Constraints on Early Acceleration\label{sec:fisher}} 

\begin{figure}
  \begin{center}{
  \includegraphics[width=\columnwidth]{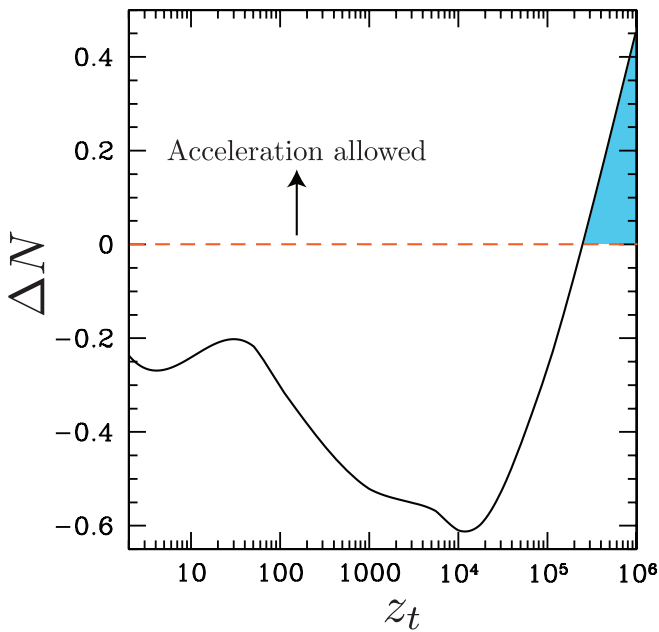}
  }
  \end{center}
  \caption{The curve gives the 95\% cl limit, assuming Planck CMB temperature
sensitivity, on $\Delta N$, the number of superdecelerating e-folds relative 
to that required for acceleration.  
Only in the upper right shaded region does the bound allow 
acceleration, $\Delta N\ge0$. 
For redshifts $z_t<2.5\times 10^5$ early acceleration is ruled out. 
}
  \label{fig:fisher}
\end{figure}

For any particular redshift in cosmic history, we can ask how 
much acceleration is permitted, i.e.\ how many e-folds $\nacc$ 
can occur starting at that time.  Carrying out a Fisher matrix 
analysis using the CMB temperature power spectrum only, with 
Planck sensitivity, we find that no extra period of acceleration is 
permitted for $z_t<2\times10^5$ at more than 99\% confidence 
level.  However the constraints rapidly weaken for higher redshifts, 
with acceleration being allowed for $z_t>3\times 10^5$. 
The results are displayed in Fig.~\ref{fig:fisher}.  
With WMAP temperature sensitivity one cannot exclude extra 
acceleration at $z_t>4\times 10^4$.  Inclusion of polarization 
data, or matter power spectra, would further tighten the bounds.  
The main point is that 
no additional period of acceleration is allowed to ameliorate the 
coincidence problem of current acceleration.

%%%%%%%%%%%%%%%%%%%%%%%%%%%%%%%%%%%%%%%%%%%%%%%%%%%%%%%%%%%%%%%
\section{Conclusions \label{sec:sumy}}

A basic question to ask about the cosmic expansion history is whether 
it is really as simple as the standard picture of a radiation dominated 
era giving way to matter domination and later the current epoch of 
acceleration.  Could there have been acceleration earlier, other than 
high energy inflation, disrupting these eras?  We demonstrate that 
the cosmic microwave background data delivers clear support of the 
standard picture, forbidding acceleration between $1\lesssim z\lesssim 
2\times 10^5$. 

The model chosen for early acceleration is the most conservative case 
in the sense that we dilute the effects of the extra energy density 
as quickly as possible ($w_j=1$), at least within the quintessence 
framework.  Even so, no solution to the coincidence problem is 
possible by taking acceleration to be an occasional phenomenon. 
Our current epoch of acceleration appears to be essentially unique 
within the last factor 100,000 of cosmic expansion, or the energy range 
$5\times 10^{-4}\,{\rm eV} \lesssim T \lesssim 25$ eV.  

Observations of the CMB are sensitive to acceleration and energy density 
through a range of physical effects, at various epochs and angular scales. 
Acceleration gives extra power to the integrated Sachs-Wolfe effect and 
the small scale photon perturbations through decay of potentials, while 
extra energy density affects geometric distance factors, shifting the 
acoustic peaks to larger or smaller scales depending on whether the 
modification is post- or pre-recombination. 

For the epochs between $z\approx 10^5-10^9$, and between primordial 
nucleosynthesis and inflation, no observational probes constrain the 
expansion history.  Detection and characterization of dark matter may 
eventually offer one window on these epochs, as can theoretical models 
of baryogenesis or the production of axions (see, e.g., \cite{grin}).  
Large swaths 
of early cosmic history remain dark, and possibly filled with dark energy.

%%%%%%%%%%%%%%%%%%%%%%%%%%%%%%%%%%%%%%%%%%%%%%%%%%%%%%%%%
\acknowledgments

We thank Ed Copeland, Marina Cort{\^e}s, Sudeep Das, Roland de Putter, 
Kim Griest, and Wayne Hu 
for useful discussions.  EL thanks the Centro de Ciencias Pedro Pascual 
in Benasque, Spain and TLS thanks the Institute for the Early 
Universe, Ewha University, Korea for hospitality.  
This work has been supported in part by the Director, Office of Science, 
Office of High Energy Physics, of the U.S.\ Department of Energy under 
Contract No.\ DE-AC02-05CH11231, and the World Class University grant 
R32-2009-000-10130-0 through the National Research Foundation, Ministry 
of Education, Science and Technology of Korea.


\begin{thebibliography}{99}

\bibitem{kaplinghat}
S.M. Carroll \& M. Kaplinghat, Phys. Rev. D 65, 063507 (2002) 
[arXiv:astro-ph/0108002]

\bibitem{zahn}
S. Galli, A. Melchiorri, G.F. Smoot, O. Zahn, Phys. Rev. D 80, 023508 (2009) 
[arXiv:0905.1808] 

\bibitem{uniq}
E.V. Linder, Phys. Rev. D 82, 063514 (2010) [arXiv:1006.4632] 

\bibitem{dodelson00}
S. Dodelson, M. Kaplinghat, E. Stewart, Phys. Rev. Lett. 85, 5276 (2000)
[arXiv:astro-ph/0002360]

\bibitem{griest}
K. Griest, Phys. Rev. D 66, 123501 (2002) [arXiv:astro-ph/0202052]

\bibitem{barenboim}
G. Barenboim, O. Mena, C. Quigg, JCAP 0604, 008 (2006) [arXiv:astro-ph/0510178]

\bibitem{lyth} 
D.H. Lyth \& E.D. Stewart, Phys. Rev. D 53, 1784 (1996) 
[arXiv:hep-ph/9510204] 

\bibitem{ackerman} 
L. Ackerman, W. Fischler, S. Kundu, N. Sivanandam, 
arXiv:1007.3511

\bibitem{liddle} 
S. Ilic, M. Kunz, A.R. Liddle, J.A. Frieman, Phys. Rev. D 81, 103502 (2010) 
[arXiv:1002.4196] 

\bibitem{sarkar} 
P. Hunt \& S. Sarkar, Phys. Rev. D 76, 123504 (2007) [arXiv:0706.2443]

\bibitem{shie}
K-F. Shie, J.M. Nester, H-J. Yo, Phys. Rev. D 78, 023522 (2008) 
[arXiv:0805.3834] 

\bibitem{linhut05}
E.V. Linder \& D. Huterer, Phys. Rev. D 72, 043509 (2005) 
[arXiv:astro-ph/0505330] 

\bibitem{hu}
W. Hu, Astrophys. J. 506, 485 (1998)
[arXiv:astro-ph/9801234] 

\bibitem{ma_and_bert}
C. P. Ma and E. Bertschinger, Astrophys. J. 455, 7 (1995)
[arXiv:astro-ph/9506072]

\bibitem{camb}
http://camb.info \\ 
A. Lewis \& A. Challinor, Astrophys. J. 538, 473 (2000) 
[arXiv:astro-ph/9911177] 

\bibitem{husugi} 
W. Hu and N. Sugiyama, Astrophys. J. 444, 489 (1995)
[arXiv:astro-ph/9407093] 

\bibitem{joneswyse}
B.J.T. Wyse \& R.F.G. Jones, Astr. Astrophys. 149, 144 (1985) 

\bibitem{lin97}
E.V. Linder, arXiv:astro-ph/9712159 

\bibitem{wmap} 
N. Jarosik et al., arXiv:1001.4744 

\bibitem{dodelson}
S. Dodelson, \emph{Modern Cosmology}, (Academic Press, 2003) 

\bibitem{act}
%J. Fowler et al., arXiv:1001.2934 
S. Das et al., arXiv:1008.0847

\bibitem{spt}
M. Lueker et al., Astrophys. J. 719, 1045 (2010) [arXiv:0912.4317] 

\bibitem{planck} 
http://planck.esa.int

\bibitem{grin}
D. Grin, T. L. Smith, M. Kamionkowski, Phys. Rev. D 77, 085020 (2008) 
[arXiv:0711.1352]


\end{thebibliography}
\end{document}